# Interference control of nonlinear excitation in a multiatom cavity QED system

Guoqing Yang[1-2], Zheng Tan[3-4], Bichen Zou[1], and Yifu Zhu[1]
[1]Department of Physics, Florida International University, Miami, Florida 33199, USA
[2]Institute of Optics, Department of Physics, Zhejiang University, Hangzhou, 310027, China
[3]Wuhan Institute of Physics and Mathematics, Chinese Academy of Sciences, Wuhan 430071, China
[4] Center for cold atom physics, Chinese Academy of Sciences, Wuhan, China 430071

We show that by manipulating quantum interference in a multi-atom cavity QED system, the nonlinear excitation of the cavity-atom polariton can be resonantly enhanced while the linear excitation is suppressed. Under appropriate conditions, it is possible to selectively enhance or suppress the polariton excitation with two free-pace laser fields. We report an experiment with cold Rb atoms in an optical cavity and present experimental results that demonstrate such interference control of the cavity QED excitation and its direct applications for studies of all-optical switching and cross-phase modulation of the cavity transmitted light.

Manipulation and control of the photon-atom interaction is a fundamental theme in optical physics. Due to the enhanced interaction of photons and atoms in a cavity, cavity quantum electrodynamics (CQED) has been developed into an important research field [1-3]. On the other hand, quantum interference such as electromagnetically induced transparency (EIT) has been used as a powerful tool to manipulate the coherent interactions of atoms and laser fields, and to enhance the optical nonlinearities, which has been widely applied in fundamental and practical studies of nonlinear optics and quantum electronics [4-5]. Naturally, exploring the atomic coherence and interference in a multi-level atomic system confined in a cavity opens new avenue to study the rich spectral and dynamical features in multi-level atomic systems modified by the CQED effects and may reveal new ways to manipulate quantum states of the coupled atoms and photons in a controlled environment [6-9].

Here we combine studies of CQED with quantum coherence and interference, and propose a method for controlling quantum states of a multi-atom CQED system. Specifically, we show that with a free-space coupling laser tuned to one of the polariton states (the normal modes) of a CQED system containing four-level atoms, destructive interference is induced in the collectively coupled cavity-atom system and the linear excitation of the cavity-atom polariton state is suppressed [10]; but with addition of a second free-space control laser, the nonlinear excitation of the polariton state is resonantly enhanced. We carried out an experiment with cold Rb atoms in an optical cavity that demonstrates the interference control of the quantum states of the CQED system. The experiment reveals the intricate features of the evolution of the quantum interference from constructive to destructive via experimentally controllable parameters, which can be used to realize the all-optical switching and cross-phase modulation of the cavity transmitted light.

We note that it has been shown that EIT in a free-space four-level system can be used to suppress the single-photon absorption and enhance the two-photon absorption [11]. The cavity EIT and its manifestation in multi-level atomic systems have been studies lately [6-9]. Recently, all optical switching in the cavity confined four-level EIT system has been demonstrated [12-13]. Although our study presented here relies on the quantum interference in a four-level atomic system, it differs from the cavity EIT in that the quantum interference here is induced in the polariton state of the CQED system and no dark state is created (the probe light cannot be coupled into the cavity) while the cavity EIT relies on the creation of an intra-cavity dark state. The difference leads to the totally different manifestation on the cavity transmitted light and will be further clarified in late discussions.

Fig. 1(a) shows the schematic diagram for the coupled CQED system that consists of N four-level atoms coupled

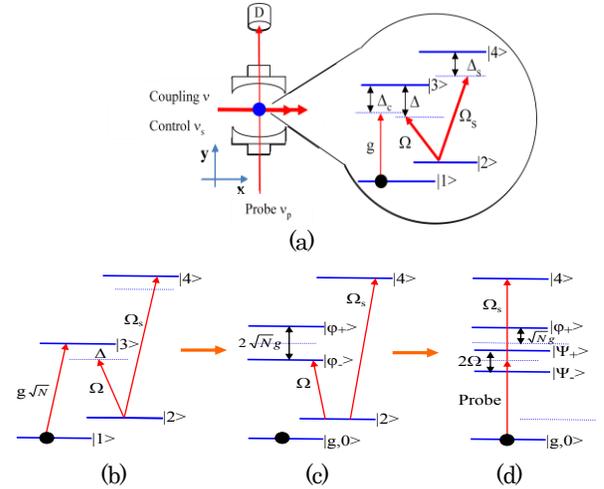

Fig.1 (a) Coherently coupled multi-atom CQED system. (b) The bare state picture with $\Delta_c=0$, $\Delta = g\sqrt{N}$, and $\Delta_s=0$. (c) The collective coupling of the N atoms and the cavity mode produces two polariton states $|\varphi_+\rangle$ and $|\varphi_-\rangle$ separated by the vacuum Rabi splitting $2g\sqrt{N}$. Also shown are the atomic states coupled by the free-space laser fields. (d) The dressed-state picture of the coherently coupled CQED system. Two excitation paths $|g,0\rangle$-$|\Psi_+\rangle$ and $|g,0\rangle$-$|\Psi_-\rangle$ are open for the intra-cavity probe light.

to a single mode cavity and also interacting with a control laser and a coupling laser from free space. The cavity mode couples the atomic transition $|1\rangle$-$|3\rangle$ with frequency detuning $\Delta_c = \nu_{cav} - \nu_{13}$. The coupling laser drives the transition $|2\rangle$-$|3\rangle$ with Rabi frequency $2\Omega$ and frequency detuning $\Delta = \nu - \nu_{23}$. The control laser drives the transition $|2\rangle - |4\rangle$ with Rabi frequency $2\Omega_s$. and frequency detuning $\Delta_s = \nu_s - \nu_{24}$. A weak probe laser is coupled into the cavity mode with frequency detuning $\Delta_p = \nu_p - \nu_{13}$. The system Hamiltonian with the collective operators $J_{31} = \sum_{i=1}^{N} \sigma_{31}^i$, $J_{32} = \sum_{i=1}^{N} \sigma_{32}^i$, and $J_{42} = \sum_{i=1}^{N} \sigma_{42}^i$. is

$H = g\hat{a}J_{31} + \Omega J_{32} + \Omega_s J_{42} + H.C.$ Here $g = \mu_{13}\sqrt{\omega_c / 2\hbar\varepsilon_0 V}$, $\hat{\sigma}_{lm}^{(i)}$ (l, m=1-4) is the atomic operator for the ith atom, and $\hat{a}$ is the annihilation operator of the cavity photons. If both $\Omega$ and $\Omega_s \ll \sqrt{N}g$, the two free-space laser fields can be treated as perturbations to the cavity coupled atomic system. When the cavity mode is tuned to the atomic transition frequency ($\Delta_c=0$), the collective coupling of the cavity mode and the atoms forms the Dicke-type atomic and photonic states [16]. The ground state of the cavity-atom system is $|g,0\rangle = |g\rangle|n=0\rangle$ ($|g\rangle = |1_1....1_i....1_N\rangle$), and the two product states with one excitation quanta are $|g,1\rangle = |g\rangle|n=1\rangle$ and $|e,0\rangle = |e\rangle|n=0\rangle$ ($|n=1\rangle$ and $|n=0\rangle$ are the number states of the cavity mode, and $|e\rangle = \sum_{i=1}^{N} 1/\sqrt{N} |1_1....3_i....1_N\rangle$ is the atomic state with only one atom in the excited state $|3\rangle$). The ground state of the CQED system is $|g,0\rangle = |g\rangle|0\rangle$, and $|\varphi_+\rangle = 1/\sqrt{2}(|e,0\rangle + |g,1\rangle)$ and $|\varphi_-\rangle = 1/\sqrt{2}(|e,0\rangle - |g,1\rangle)$ are two excited eigenstates separated in frequency by the vacuum Rabi splitting $2\sqrt{N}g$ (see Fig. 1(c)) and are referred to as the cavity-atom polariton states (the normal modes) [14-16]. When the coupling field is tuned to be resonant with one of the polariton states $|\varphi_-\rangle$ as shown in Fig. 1(c), the polariton state $|\varphi_-\rangle$ is split into two dressed polariton states, $|\Psi_+\rangle = 1/\sqrt{2}(|\varphi_-\rangle + |2'\rangle)$ an $|\Psi_-\rangle = 1/\sqrt{2}(|\varphi_-\rangle - |2'\rangle)$, ($|2'\rangle = \sum_{i=1}^{N} 1/\sqrt{N}|1_1....2_i....1_N\rangle$), which are separated in frequency by $2\Omega$. With the probe laser coupled into the cavity and its frequency resonant with the polariton $|\varphi_-\rangle$, two excitation paths, $|g,0\rangle \to |\Psi_+\rangle$ and $|g,0\rangle \to |\Psi_-\rangle$ interfere destructively, which leads to the suppression of the linear excitation of the polariton $|\varphi_-\rangle$ and blocks the probe light transmission through the cavity. When the free-space control laser is present, it disrupts the destructive interference, restores the polariton excitation and enables the probe light to transmit through the cavity.

It is interesting to compare the CQED system studied here with the four-level cavity EIT system reported earlier [17]. In the cavity EIT system, the probe light transmitted through the cavity is peaked at $\Delta=\Delta_p=0$ without the control laser (the atom-cavity system is transparent to the probe light) [9-13]; with the control laser, the intra-cavity dark state splits into two dressed dark states (the cavity EIT peak splits into two peaks). Here with suppression of the polariton excitation from the coupling induced interference in the CQED system, the probe light transmission at $\Delta_p = \Delta = \pm g\sqrt{N}$ is blocked without the control laser (the CQED system is opaque to the probe light); with the control laser present, the destructive interference is disrupted and the polariton excitation is restored, enabling the probe light to transmit through the cavity.

The above qualitative physical picture is confirmed by the quantitative analysis. The equations of motion for the CQED system, $\frac{d\hat{\rho}}{dt} = -[H,\hat{\rho}] + \hat{L}\hat{\rho}$, can be derived from the cavity input-output theorem. With a weak probe field, we obtain the steady-state solution of the intra-cavity probe field. Under the polariton resonance conditions $\Delta = \Delta_p = \pm g\sqrt{N}$ and $\Delta_s=0$, the transmitted probe field is

$$a_T = \frac{\kappa a_p^{in}}{\kappa \mp ig\sqrt{N} + \frac{g^2 N(\Omega_s^2 + (\gamma_{21}+\Gamma_4)\gamma_{21})}{(\Gamma_3 \mp ig\sqrt{N})(\Omega_s^2 + (\Gamma_4+\gamma_{21})\gamma_{21}) + \Omega^2(\Gamma_4+\gamma_{21})}} \quad (1)$$

Here $\Gamma_3$ ($\Gamma_4$) is the decay rate of the excited state $|3\rangle$ ($|4\rangle$), $\gamma_{21}$ is the decoherence rate between the ground states $|1\rangle$ and $|2\rangle$, and $\kappa$ is the cavity decay rate. The transmitted probe field is proportional to the photonic part of the polariton state $|\varphi_+\rangle$ (or $|\varphi_-\rangle$) and therefore represents the excitation of $|\varphi_\pm\rangle$ by the probe light. To reveal the essential physics, we consider the regime of strong collective cavity-atom coupling $g\sqrt{N} \gg \Gamma_3, \Gamma_4, \kappa, \Omega$, and $\Omega_s$, and neglect the ground state decoherence ($\gamma_{21}=0$). Then Eq.(1) shows that (a) when both $\Omega_s=0$ and $\Omega=0$, the cavity transmitted probe field is $a_T = \kappa a_p^{in}/(\kappa+\Gamma_3)$, which represents the single-photon excitation of the cavity-atom polariton state; (b) with $\Omega \neq 0$ (the coupling laser is present) and $\Omega_s=0$ (no control laser), the cavity transmitted probe field is $a_T = 0$. This shows that the coupling field induces destructive quantum interference that suppresses the excitation of the cavity-atom polariton and blocks the probe light transmission through the CQED system; (c) when both the coupling laser and the control laser are present ($\Omega \neq 0$ and $\Omega_s \neq 0$), the cavity transmitted probe field becomes $a_T = \kappa\Omega_s^2 a_p^{in}/(\Omega_s^2(\Gamma_3+\kappa) + \Gamma_4\Omega^2)$, which shows that the transmitted probe field is proportional to $\Omega_s^2$ (when $\Omega \gg \Omega_s$) and the control laser results in the nonlinear excitation of the cavity-atom polariton state (see Fig. 1(d)). For a large transmission of the probe light, it is desirable to have a cavity with a sufficiently large decay rate $\kappa$. Since the identical results are obtained for $\Delta = \Delta_p = g\sqrt{N}$ or $\Delta = \Delta_p = -g\sqrt{N}$, the further analysis and experimental data in the following sections will be presented for the polariton state $|\varphi\rangle$ with $\Delta = -g\sqrt{N}$ and $\Delta_p$ near $-g\sqrt{N}$.

It is instructive to see how the transmitted probe light changes with the probe frequency and the control frequency when the coupling laser is set at the polariton resonance $\Delta = -g\sqrt{N}$ and induces the destructive interference for the linear polariton excitation. Fig. 2 plots the calculated intensity of the cavity-transmitted probe light $I_t$ (normalized by the input intensity $I_{in}$) versus the probe detuning $\Delta_p/\Gamma_3$ and the control detuning $\Delta_s/\Gamma_3$ across the polariton state $|\varphi_-\rangle$ ($\Delta = -g\sqrt{N}$). When the control detuning $\Delta_S$ is large, its effect can be neglected and the polariton excitation is suppressed at the probe detuning $\Delta_p = -g\sqrt{N}$ (=$-10\Gamma_3$ in Fig 2). As the control laser is tuned near the resonance $\Delta_S=0$, the central dip at $\Delta_p=-g\sqrt{N}$ evolves into a peak, representing the resonantly enhanced nonlinear excitation of the polariton state (the two outside peaks represent the linear excitation of the two dressed polariton states $|\Psi_+\rangle$ and $|\Psi_-\rangle$, see Fig. 1(d)).

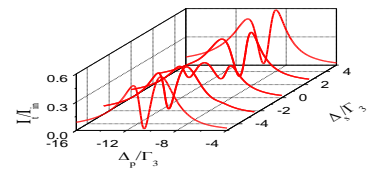

Fig. 2 Cavity-transmitted probe intensity $I_t/I_{in}$ versus $\Delta_p/\Gamma_3$ and $\Delta_s/\Gamma_3$. The parameters are $\Delta_c=0$, $\Gamma_4=\Gamma_3$, $g\sqrt{N} = 10\Gamma_3$, $\Delta = -g\sqrt{N}$, $\kappa=3\Gamma_3$, $\gamma_{12}=0.01\Gamma_3$, and $\Omega=\Omega_s=1.5\Gamma_3$.

The experiment is done with cold $^{85}$Rb atoms in a magneto-optical trap (MOT). A tapered-amplifier diode laser with output power ~300 mW is used as the cooling laser. An extended-cavity diode laser with an output power of ~20 mW is used as the repump laser. The beam diameter of both lasers is ~ 1 cm. The cavity consists of two mirrors of 5 cm curvature separated by ~ 5 cm and is mounted on an Invar holder inside the vacuum chamber. The empty cavity finesse is measured to be ~ 35. The trapped $^{85}$Rb atom cloud in the MOT is ~ 2 mm with about 5x10$^5$ atoms inside the effective mode volume of the cavity. The four-level atomic system of Fig. 1 is realized with the $^{85}$Rb D$_1$ transitions, in which the F=2 and F=3 ground states are chosen as the states |1> and |2>, and the excited states F'=3 and F'=2 are chosen as the states |3> and |4>, respectively. Three extended-cavity diode lasers at 795 nm are used as the probe laser (couples the F=2-F'=3 transitions), the coupling laser (drives the F=3-F'=3 transitions) and the control laser (couples the F=3-F'=2 transitions). We define the quantization axis in the x direction (perpendicular to the cavity axis, see Fig.1a). The coupling laser is σ+ polarized and the control laser is π polarized in the y direction. The two lasers have a beam diameter of ~ 5 mm, and are made to co-propagate perpendicularly to the intra-cavity probe beam to intercept the cold Rb atoms at the cavity center. The attenuated probe beam is π polarized parallel to the x direction and is coupled into the cavity through a mode-matching lens. The cavity-transmitted probe light passes through an iris and is coupled into a multi-mode fiber, the output of which is collected by a photon counter.

The experiment is run at a repetition rate of 10 Hz. All lasers are turned on or off by acousto-optic modulators (AOM). For each period of 100 ms, ~98.9 ms is used for cooling and trapping of the $^{85}$Rb atoms, during which the trapping laser and the repump laser are turned on by two AOMs while the coupling laser, the control laser, and the probe laser are off. The time for the data collection lasts ~ 1.1 ms, during which the cooling laser, the repump laser, and the DC current to the MOT anti-Helmholtz coils are turned off, and the coupling laser, the control laser, and the probe laser are turned on by separate AOMs.

Fig. 3 plots the cavity transmitted intensity of the probe light versus the probe detuning $\Delta_p$. The values of $g\sqrt{N}$, κ, Ω, and Ω$_S$ are determined from the measured vacuum Rabi frequency, the empty cavity linewidth, and the laser intensities respectively, and then are used in the numerical calculations. Fig. 3(a) shows the probe transmission spectrum without the coupling laser and the control laser, and exhibits two peaks representing the two polariton states $|\varphi_+\rangle$ and $|\varphi_-\rangle$. Fig. 3(b) shows the probe transmission spectrum when the coupling laser with $\Delta = -g\sqrt{N}$ is present but without control laser (Ω$_S$=0). The coupling laser induces the destructive interference that suppresses the polariton excitation and leads to a narrow dip at $\Delta_p = -g\sqrt{N}$. Fig. 3(c) shows that when the resonant control laser is added, three peaks appear around $\Delta_p = -g\sqrt{N}$: the central peak at $\Delta_p = -g\sqrt{N}$ represents the enhanced nonlinear excitation of the polariton while the two side peaks at $\Delta_p = -g\sqrt{N} \pm \Omega$ represent the linear excitation of the dressed polariton states $|\Psi_+\rangle$ and $|\Psi_-\rangle$ (see Fig. 1(d)).

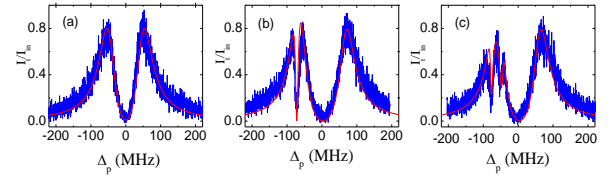

Fig. 3 Transmitted probe intensity versus the probe frequency detuning $\Delta_p$. (a) The coupling laser and the control laser are absent (Ω=Ω$_S$=0). (b) The coupling laser is on and tuned to the polariton resonance $\Delta = -g\sqrt{N}$ while the control laser is off (Ω$_S$=0). (c) Both the coupling laser and the control laser are on (with $\Delta = -g\sqrt{N}$ and $\Delta_S = 0$). Blue lines are the experimental measurements and red lines are the numerical calculations. The parameters used in the calculations are $\Delta_c = 0$, Ω=Ω$_S$=15 MHz, $g\sqrt{N}$ = 66 MHz, κ=48 MHz, γ$_{12}$=0.01Γ$_3$.

Fig. 4 plots the measured transmitted probe intensity I$_t$/I$_{in}$ versus the control laser intensity I$_s$. It shows the nonlinear nature of the control-enabled polariton excitation. I$_t$/I$_{in}$ increases rapidly versus the control laser intensity and saturates at high control intensities.

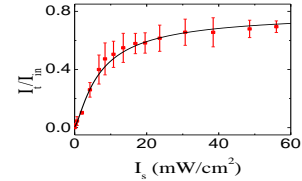

Fig. 4 Ratio of the transmitted probe intensity I$_t$ and the input probe intensity I$_{in}$ versus the control laser intensity I$_s$. The solid line is the theoretical calculation with $\Delta = \Delta_p = -g\sqrt{N}$ =-66 MHz (the other parameters are the same as those in Fig. 3).

By varying the probe detuning $\Delta_p$ around the polariton resonance, a multi-facet behavior was observed: the control laser can be made to enhance or suppress the polariton excitation. Such interference control of the polariton excitation can be used for studies of light controlling light as shown in Fig. 5. The top panel of Fig. 5 plots I$_t$/I$_{in}$ versus the control detuning $\Delta_S$ (with $\Delta_p$ set at fixed values and $\Delta = -g\sqrt{N}$). The bottom panel of Fig. 5 plots the calculated phase shift of the transmitted probe light from the cross-phase modulation of the control light. If there is no control laser, the transmitted probe light is peaked at $\Delta_p=-g\sqrt{N}-\Omega$ due to the resonant linear excitation of the dressed polariton states $|\Psi_-\rangle$, (see Fig. 1(d) and Fig. 3(b)). Then with the probe frequency set at $\Delta_p=-g\sqrt{N}-\Omega$ and the control laser on, the transmitted probe intensity versus $\Delta_S$ is plotted in Fig. 5(a), which exhibits a dip at $\Delta_S \approx \Omega$. This dip is due to the two-photon transition |g,0>- |$\Psi_-$> - |4> induced by the intra-cavity

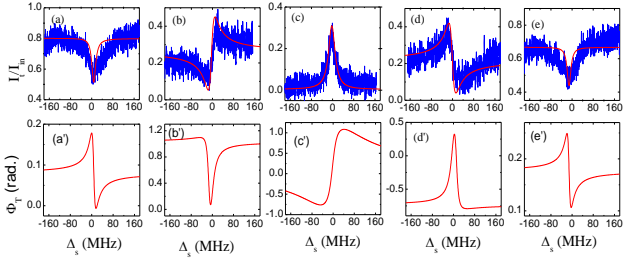

Fig. 5 (a) to (e), transmitted probe intensity versus the control frequency detuning $\Delta_S$. (a) $\Delta_p \approx$ -76 MHz; (b) $\Delta_p \approx$ -71 MHz; (c) $\Delta_p \approx$ -66 MHz; (d) $\Delta_p \approx$ -60 MHz; and (e) $\Delta_p \approx$ -56 MHz. Blue lines are experimental measurements and red lines are theoretical calculations. Fig. (a') to (e'), calculated phase shift $\Phi_t$ of the transmitted probe light versus $\Delta_S$. The coupling detuning $\Delta = -g\sqrt{N}$ and $\Omega \approx$ 10 MHz. The other parameters are $\Delta_c = 0$, $\Omega_S$=6 MHz, $g\sqrt{N} = 66$ MHz, $\kappa$=48 MHz, $\gamma_{12}$=0.01$\Gamma_3$.

probe field and the free-space control field. The two-photon transition leads to the excitation of the state |4⟩ and results in a decrease of the intra-cavity probe photons. Consequently, the transmitted probe light is reduced at the two-photon resonance $\Delta_S = \Omega$. The corresponding phase shift $\Phi_t$ of the transmitted probe light exhibits an anomalous dispersive lineshape (Fig. 5(a')). Similarly, when the probe frequency is tuned to the resonance of the other dressed polariton state |Ψ+⟩ (Fig. 5(e) at $\Delta_p = -g\sqrt{N} + \Omega$), the two-photon transition |g,0⟩-|Ψ+⟩-|4⟩ induced by the control field and the intra-cavity probe field is resonant at $\Delta_S$=-$\Omega$, and results in a dip for the transmitted probe light (Fig. 5(e)) and an anomalous dispersive lineshape for $\Phi_t$ (Fig. 5(e')). When the probe laser is tuned to the resonance of the polariton states |φ⟩ (Fig. 5(c) at $\Delta_p$=-$g\sqrt{N}$), the linear excitation of the polariton state is suppressed by the coupling-laser induced interference, but the control laser disrupts the coupling-induced destructive interference and restores the polariton excitation. The spectral line profile of the transmitted probe light exhibits a peak at $\Delta_S$=0. The phase shift $\Phi_t$ exhibits a normal dispersive lineshape (Fig. 5(c')). When the probe frequency is set between the dressed polariton resonance at $\Delta_p = -g\sqrt{N} \pm \Omega$ and the polariton resonance $\Delta_p = -g\sqrt{N}$, the probe transmission versus $\Delta_S$ exhibits a dispersive lineshape, and either a normal (Fig. 5(b), $\Delta_p$ is red detuned from the polariton resonance) or anomalous dispersive lineshape (Fig. 5(d), $\Delta_p$ is blue detuned from the polariton resonance) is observed. Correspondingly, the phase shift $\Phi_t$ exhibits either a dip (Fig. 5(b')) or a peak (Fig. 5(d')).

Fig. 5 shows that the interference controlled polariton excitation in the CQED system can be used for studies of all-optical switching and cross-phase modulation of the cavity-transmitted probe light. The system parameters for such studies can be identified from Fig. 5. For the all-optical-switching study, the system should be set at $\Delta$=$\Delta_p$=-$g\sqrt{N}$ and $\Delta_S$=0 (Fig. 5(c)). Then, when the control laser is off, the transmitted probe light is near zero (the switch is open); when the control laser is on, the transmitted probe light is peaked (the switch is closed), which provides a high switching contrast. For the cross-phase modulation, it is desirable to produce a large phase shift while maintaining a sufficiently large intensity for the transmitted probe light. This requirement is met by the parameters of Fig. 5(b) and 5(d), in which $\Delta = -g\sqrt{N}$, $\Delta_S$=0, and $\Delta_p$ is set to the value between the dressed polariton resonance at $\Delta_p = -g\sqrt{N} \pm \Omega$ and the polariton resonance $\Delta_p = -g\sqrt{N}$. Under such conditions, the phase shift can be inferred from the measured intensity of the transmitted probe light in our experiments. Specifically, the peak phase shift is -0.98 rad. (with It/I$_{in}$≈30%) in Fig. 5(b') and 0.99 rad. (with It/I$_{in}$≈20%) in Fig. (5d').

In conclusion, we have shown that a free-space coupling laser can be used to induce destructive interference that suppresses the linear polariton excitation in a CQED system. The coupling induced interference can be disrupted by adding a free-space control laser, which leads to the enhanced nonlinear excitation of the cavity QED system. We carried out experiments in a cavity-atom system and demonstrated the interference control of the polariton state excitation. Our experiment shows the interplay of the CQED and the quantum interference, and opens new ways to study all-optical switching and cross-phase modulation in a multi-atom CQED system.

### Acknowledgement
This work is supported by the National Science Foundation under Grant No. 1205565.